\documentclass[11pt,a4paper]{article}
\pdfoutput=1
\usepackage[hyperref]{acl2019}
\usepackage{times}
\usepackage{latexsym} 

\usepackage[utf8]{inputenc} 
\usepackage[T1]{fontenc}    
\usepackage{hyperref}       
\usepackage{url}            
\usepackage{booktabs}       
\usepackage{amsfonts}       
\usepackage{nicefrac}       
\usepackage{microtype}      
\usepackage[tableposition=top]{caption}
\usepackage{lipsum}
\usepackage{graphicx}
\usepackage{subfig}

\aclfinalcopy

\title{``The Michael Jordan of Greatness'':\\Extracting Vossian Antonomasia from Two Decades of\\ the \emph{New York Times}, 1987--2007\Thanks{This is a preprint. The final published version of this article is available at DOI:\href{https://doi.org/10.1093/llc/fqy087}{10.1093/llc/fqy087}.}}

\author{
  Frank Fischer\\
  National Research University\\Higher School of Economics, Moscow\\
  \href{mailto:ffischer@hse.ru}{ffischer@hse.ru} \\
  \And
  Robert Jäschke\\
  Humboldt-Universität zu Berlin\\
  \href{mailto:robert.jaeschke@hu-berlin.de}{robert.jaeschke@hu-berlin.de} \\
}

\newcommand{\zeitung}[2]{(in \emph{#1}, #2)}

\begin{document}
\maketitle

\begin{abstract}
Vossian Antonomasia is a prolific stylistic device, in use since antiquity. It can compress the introduction or description of a person or another named entity into a terse, poignant formulation and can best be explained by an example: When Norwegian world champion Magnus Carlsen is described as ``the Mozart of chess'', it is Vossian Antonomasia we are dealing with. The pattern is simple: A source (Mozart) is used to describe a target (Magnus Carlsen), the transfer of meaning is reached via a modifier (``of chess''). This phenomenon has been discussed before (as `metaphorical antonomasia' or, with special focus on the source object, as `paragons'), but no corpus-based approach has been undertaken as yet to explore its breadth and variety. We are looking into a full-text newspaper corpus (\emph{The New York Times}, 1987--2007) and describe a new method for the automatic extraction of Vossian Antonomasia based on Wikidata entities. Our analysis offers new insights into the occurrence of popular paragons and their distribution.
\end{abstract}

\section{Introduction and Related Works}

When Peter Paul Rubens is described as ``the Tarantino of Baroque painting'' \zeitung{Tagesspiegel}{2014} or Slavoj Žižek as ``the Elvis of cultural theory'' \zeitung{Slate}{2014}, we deal with a phenomenon called Vossian Antonomasia (VA). This trope is named after Gerardus Vossius (1577--1649), the Dutch humanist and professor of rhetoric, who first described VA as a specific sub-phenomenon of antonomasia. Generally speaking, an antonomasia incorporates the idea of a particular property of a person standing in for this person (e.\,g., ``The Voice'' standing in for Frank Sinatra). In the case of \emph{Vossian} Antonomasia, a person is attributed a specific property by reference to another (more well-known, more popular, more notorious) person. An ``atypological, modifying signal (pronoun, adjective, genitive)'' provides for the transfer of meaning \citep{lausberg1960handbuch}. ``Baroque painting'' and ``cultural theory'' work as such in the above-mentioned examples. In other words, VA uses a 'modifier' to establish a relationship between a `source' and a `target' \citep{bergien2013names}. In our paper, we will study this popular variant, although VA generally works without such modifier (as in ``Go and denounce me, you Judas!'', where Judas stands for "traitor" and no modifier is presented). Entities can occur both as source and as target, as is shown by the example of Obama in \citet{bergien2013names}: until 2011 he mostly appeared as target, but increasingly as source thereafter. Following \citet{lakoff1987women}, the entity serving as `source' is also referred to as `paragon', ``a specific example that comes close to embodying the qualities of the ideal'' \citep{bergien2013names}.

Internationally, the term `Vossian Antonomasia' is rarely used; instead, we encounter a distinction between `Antonomasia\textsubscript{1}' and `Antonomasia\textsubscript{2}': `metonymic' vs. `metaphorical antonomasia' \citep{holmqvist2010princess}. In this classification scheme, Vossian Antonomasia would occur within `Antonomasia\textsubscript{2}', in the form of ``comparisons with paragons from other spheres of culture: Lyotard is a pope of postmodernism, Bush is no Demosthenes; and we can buy the Cadillac of vacuum cleaners.'' \citep{holmqvist2010princess}

The use of this stylistic device dates back to classical antiquity. Nowadays, it is a welcome element in many journalistic genres and often occurs in headings, since it can be both informative and enigmatic, and is notorious for its unique entertaining qualities. A larger collection of examples\footnote{See \url{https://www.umblaetterer.de/datenzentrum/vossianische-antonomasien.html} and \citet{fischer2014napoleon}.} gave the impetus to research this phenomenon more systematically, with historical perspective and on the basis of larger English and German newspaper corpora \citep{jaeschke2017nutella}. The aim of this paper is an exploratory automated analysis of VA in the \emph{New York Times} (1987--2007). The object of study was chosen because we were looking for a corpus of a general contemporary English-language newspaper and found the NYT Annotated Corpus a perfect fit being readily available as it was released to the research community \citep{sandhaus2008nyt}.

\section{Extraction and Evaluation}

The method described in \citet{jaeschke2017nutella} was based on NLTK's NER module and complex hand-crafted regular expressions. This resulted in low precision (few of the resulting candidates were VA) and recall (only few VA were detected). Reason enough to work on a novel method for the extraction of VA. We did so by restricting our search to the clear-cut pattern ``the SOURCE of''. This pattern limitation lead to the extraction of a much larger set of VA than before (by a factor of 10) with just minimal additional effort. It has to be noted that this strategy works well for our English-language corpus, but would have to be adjusted for languages that do not use articles or prepositions to determine cases.

To overcome the shortcomings of current NER implementations we resorted to Wikidata, the crowd-sourced fact database \citep{vrandecic2014wikidata}, from which we extracted a list of possible sources of a VA expression. In most other use-case scenarios, using Wikidata for named-entity matching would lead to a deterioration or limitation of the results. But in our very special case, this simple idea was the breakthrough to achieve clean extractions and to immensely increase precision. Because one prerequisite for being a source of a VA is to be famous or popular or notorious for something, and this arguably goes along with being present on at least one of the hundreds of Wikipedia versions and thus having a Wikidata entry.

By help of the Wikidata Toolkit\footnote{See \url{https://www.mediawiki.org/wiki/Wikidata_Toolkit}.} we downloaded and filtered the Wikidata dump from August 14, 2017. We used the instanceOf\footnote{See \url{https://www.wikidata.org/wiki/Property:P31}.} property to retrieve all instances of the class `human'.\footnote{See \url{https://www.wikidata.org/wiki/Q5}.} For each individual we retrieved the English representation of their name and potential alias names. The resulting list comprises 2,801,931 individuals with 2,955,761 unique names.

From the XML-encoded \emph{New York Times} corpus \citep{sandhaus2008nyt}, we extracted the full text for each of the 1,854,726 articles and split it into sentences using the Natural Language Toolkit, NLTK \citep{bird2009nltk}. Within each sentence we then searched for character sequences matching the regular expression
{\small \verb/(\\bthe\\s+([\\w.,'-]+\\s+){1,5}?of\\b)/}.

It matches all phrases beginning with ``the'' and ending with ``of'', with up to five words in-between. We had to increase this margin to be able to include dots, commas, apostrophes and hyphens, allowing us to match sources like ``Robert Downey, Jr.'', ``Shaquille O'Neal'', or ``Jean-Luc Godard''. Each sentence containing a matched phrase was stored together with the extracted source candidate.

We then tried to match each source candidate against the Wikidata list of individuals. This effectively reduced the number of candidate sentences from 11,452,615 to 96,731, but still included one-word source candidates like ``Church'' or ``Hall'', caused by ambiguous nicknames or aliases of some persons stored in Wikidata,\footnote{E.\,g., \url{https://www.wikidata.org/wiki/Q21508608}.} or incomplete data.\footnote{ E.\,g., missing first names, see \url{https://www.wikidata.org/wiki/Q5116540}.} We decided to filter out such cases with a curated blacklist, which mainly concerned one-word aliases (of the 1,289 entries in the blacklist, only 62 consist of more than one word).\footnote{The blacklist and all other data can be found in our working repository at \url{https://github.com/weltliteratur/vossanto} and via the project website at \url{https://vossanto.weltliteratur.net/}.} After filtering the source candidates by help of our blacklist, we were down to 3,753 sentences from 20 years of \emph{New York Times} containing possible VA.

While we cannot provide numbers for recall due to missing gold annotations for the VA phenomenon, we can present numbers for the precision of our extraction method. Compared to other stylistic devices, VA are a notably rare phenomenon -- we were able to evaluate a candidate list of this magnitude manually.

\begin{figure*}
    \subfloat[Absolute numbers\label{fig:precision-abs}]{\includegraphics[width=.49\linewidth]{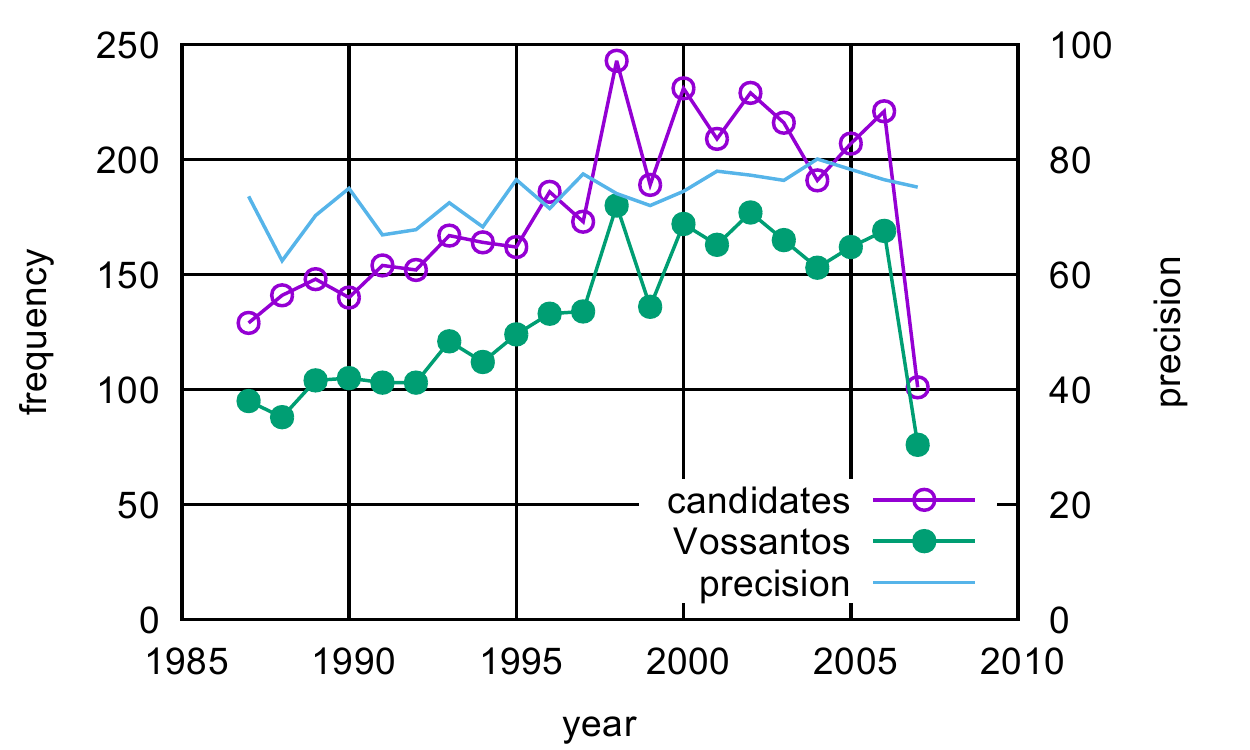}}
    \hfill
    \subfloat[Relative frequency (one per thousand)\label{fig:precision-rel}]{\includegraphics[width=.49\linewidth]{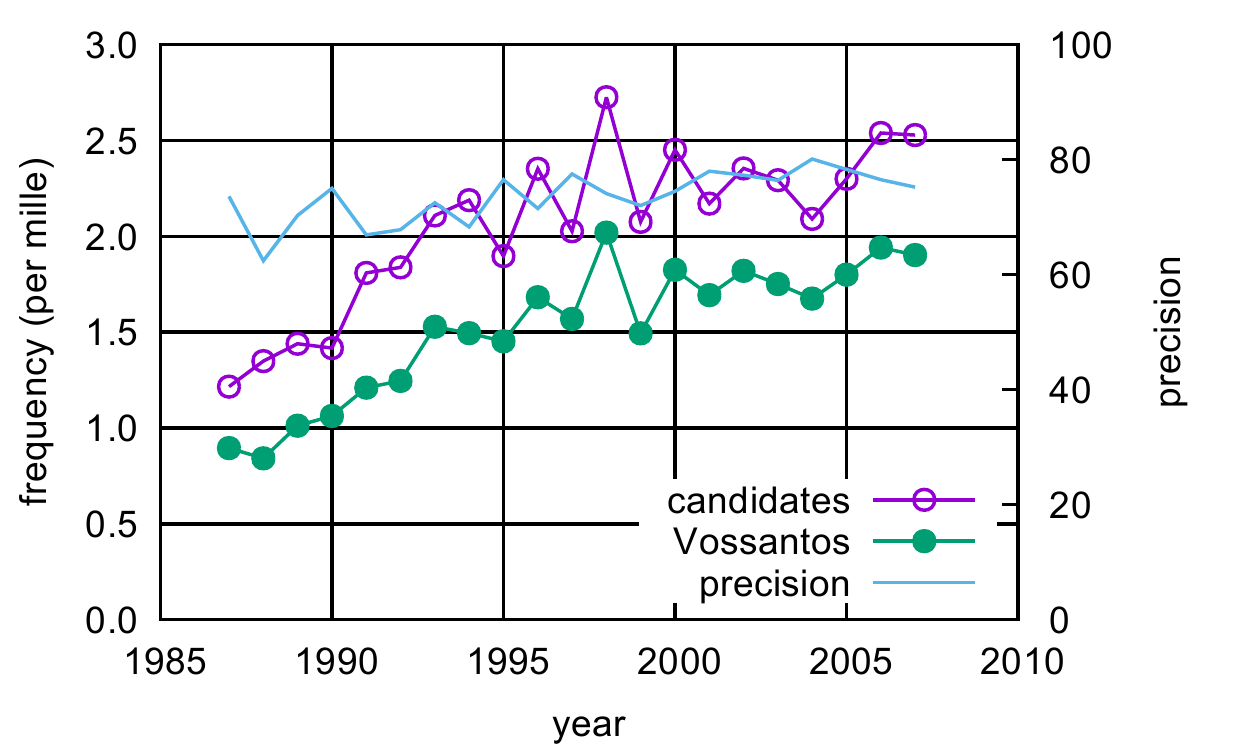}}
    \caption{VA candidates and true VA extracted per year. The right vertical axes measure the resulting precision (percentage of true VA among the candidates).}
    \label{fig:my_label}
\end{figure*}

The final result is a list of 2,775 sentences containing VA, making for a 73.9\% precision score. (While these are all cases of valid VA, we have eliminated duplicates for the rankings in Section~\ref{sec:results} if an expression reoccurred in a text or if an article was republished, trimming the list to 2,646 unique occurrences of VA.) Figure~\ref{fig:precision-abs} shows these numbers on a per-year basis (the corpus ends early in 2007, which explains the abrupt decrease in that year). The slight increase of the curves in Figure~\ref{fig:precision-rel} (relative frequency) suggests that the use of VA increased over the years covered by our corpus. Since VA are typically built around a popular person, a paragon, it is unlikely that Wikipedia/Wikidata has significant gaps for the time covered by our corpus, but we have to at least consider that Wikipedia, and as a consequence, Wikidata, too, might have a contemporary bias.

\section{Results}\label{sec:results}

\subsection{Most Frequent Sources}

The 40 most frequent sources are shown in Table~\ref{tab:sources}. They include artists, people from sports, business, the culture industry, historical figures, and scientists. While especially the upper segment of the table is dominated by (US-American) men, Madonna is the most frequent female paragon in our corpus, followed by Martha Stewart, Greta Garbo, Mother Teresa, Oprah Winfrey, Rosa Parks, Susan Lucci, and Julia Child.

\begin{table}[]
    \centering
    \caption{The forty most frequent VA sources extracted from our corpus (`the \dots of').}
    \begin{tabular}{@{}rl@{}}
    \toprule
    count & source \\
    \midrule
    68&  Michael Jordan \\          
    58&  Rodney Dangerfield \\      
    36&  Babe Ruth \\               
    32&  Elvis Presley \\           
    31&  Johnny Appleseed \\        
    23&  Bill Gates \\              
    21&  Pablo Picasso \\           
    21&  Michelangelo \\            
    21&  Donald Trump \\            
    21&  Jackie Robinson \\         
    21&  Madonna \\                 
    20&  P. T. Barnum \\            
    20&  Tiger Woods \\             
    18&  Martha Stewart \\          
    16&  Henry Ford \\              
    16&  William Shakespeare \\     
    16&  Wolfgang Amadeus Mozart \\ 
    15&  Adolf Hitler \\            
    14&  Greta Garbo \\             
    14&  John Wayne \\              
    14&  Mother Teresa \\           
    13&  Napoleon \\                
    13&  Ralph Nader \\             
    12&  Leonardo da Vinci \\       
    12&  Cal Ripken \\              
    12&  Leo Tolstoy \\             
    12&  Oprah Winfrey \\           
    12&  Rosa Parks \\              
    12&  Susan Lucci \\             
    11&  Walt Disney \\             
    11&  Willie Horton \\           
    11&  Rembrandt \\               
    10&  Albert Einstein \\         
    10&  Thomas Edison \\           
    10&  Mike Tyson \\              
    10&  Julia Child \\             
     9&  Ross Perot \\              
     9&  Dennis Rodman \\           
     8&  James Dean \\              
     8&  Mikhail Gorbachev \\
     \bottomrule
    \end{tabular}
    \label{tab:sources}
\end{table}

The top source, Michael Jordan, is followed by Rodney Dangerfield, a stand-up comedian preeminently known for his catchphrase ``I don't get no respect!''. He thus finally gets some respect, at least as a highly productive source for VA. Next in the ranking are Babe Ruth, an American baseball player, and Elvis Presley. Altogether, Table~\ref{tab:sources} assembles international celebrities, some of which are prone to appear beyond the \emph{New York Times}. Yet, many references are not likely as sources for VA in other nations' newspapers (just take all the baseball references), which confirms the notion that VA is a language- and culture-dependent phenomenon \citep{koevecses2005metaphor,jaeschke2017nutella}.

\subsection{High-Frequent Modifiers}

Looking at the most frequent modifiers, we find proof for one of the main functions of VA, which is inculturation \citep{holmqvist2010princess}. Next to unspecific modifiers such as ``of his [or her] day'', ``of his [or her] time'', etc., we can find toponyms such as Japan, China, and Brazil (Table~\ref{tab:modifier}). This becomes even clearer if we limit the output to just countries (Table~\ref{tab:modifier_countries}).

The same can be said about other kinds of modifiers, such as sports (baseball, tennis, basketball, hockey, football). They can transfer meaning between different disciplines, like in this sentence: ``It's one thing to win a gold medal, but the way he [Andre Agassi] dominated, he was the Michael Johnson of the tennis field today.'' (NYT, 1996) But given the right source, they can also inherit meaning from other domains, as shows this example: ``[MBL pitcher Greg] Maddux is the Picasso of baseball.'' (NYT, 2001)

\begin{table}[]
    \centering
    \caption{Most frequent modifiers.}
    \begin{tabular}{@{}rl@{}}
    \toprule
    count & modifier \\
    \midrule
    55 & his day\\
    33 & his time\\
    29 & Japan\\
    16 & tennis\\
    16 & his generation\\
    16 & baseball\\
    15 & China\\
    13 & her time\\
    13 & her day\\
    12 & our time\\
    11 & the 1990's\\
    10 & the Zulus\\
    10 & the 90's\\
    10 & politics\\
    10 & hockey\\
    10 & Brazil\\
    10 & basketball\\
    10 & ballet\\
     9 & jazz\\
     9 & fashion\\
     8 & today\\
     8 & Israel\\
     8 & his era\\
     8 & hip-hop\\
     8 & golf\\
     8 & dance\\
     7 & the art world\\
     7 & the 19th century\\
     7 & Long Island\\
     7 & Iran\\    
     \bottomrule
     \end{tabular}
    \label{tab:modifier}
\end{table}

\begin{table}[]
    \centering
    \caption{Selected countries serving as modifiers.}
    \begin{tabular}{@{}rl@{}}
    \toprule
    count & country \\
    \midrule
    29 & Japan\\
    15 & China\\
    10 & Brazil\\
     8 & Israel\\
     7 & Iran\\
     7 & India\\
     4 & South Africa\\
     4 & Mexico\\
     3 & Spain\\
     3 & South Korea\\
     3 & Russia\\
     3 & Poland\\
     3 & Pakistan\\
     \bottomrule
     \end{tabular}
    \label{tab:modifier_countries}
\end{table}

\subsection{Distribution of VA}
We also gained insights into the distribution of VA across our corpus. VA as a phenomenon is particularly popular in the cultural pages (``Arts'', ``Books'', ``Movies'', ``Theater'', ``Magazine'') and the ``Sports'' section, as indicated in Table~\ref{tab:newspaper_section}. In contrast to that, authors in the ``Business'' section of the \emph{New York Times} (the section containing the most articles in our corpus, see Table~\ref{tab:newspaper_categories}) clearly think twice before operating a VA expression into their texts. The popularity of VA especially on the cultural pages is confirmed by the example of German weekly \emph{Die Zeit} studied in \cite{jaeschke2017nutella} where the paper's ``Feuilleton'' section exhibits the highest frequency of VA. Our quantitative approach also backs up findings of qualitative research on antonomasia in the ``Sports'' section of Croatian media \citep{grgic2013cowboys}.

\begin{table*}[]
    \centering
        \caption{Occurrences of VA by newspaper section. Categories are as extracted from corpus, this includes empty category fields.}
    \begin{tabular}{@{}rrlrr@{}}
    \toprule
        VA &  2,646 & category               & articles & 1,854,726 \\
   \midrule
       336 & 12.7\% & Sports                 &   160,888 &    8.7\% \\
       334 & 12.6\% & Arts                   &    88,460 &    4.8\% \\
       290 & 11.0\% & New York and Region    &   221,897 &   12.0\% \\
       237 &  9.0\% & Arts; Books            &    35,475 &    1.9\% \\
       158 &  6.0\% & Movies; Arts           &    27,759 &    1.5\% \\
       109 &  4.1\% & Business               &   291,982 &   15.7\% \\
       102 &  3.9\% & Opinion                &   134,428 &    7.2\% \\
        96 &  3.6\% & U.S.                   &    89,389 &    4.8\% \\
        95 &  3.6\% & Magazine               &    11,464 &    0.6\% \\
        62 &  2.3\% & Style                  &    65,071 &    3.5\% \\
        61 &  2.3\% & Arts; Theater          &    13,283 &    0.7\% \\
        46 &  1.7\% & World                  &    79,786 &    4.3\% \\
        39 &  1.5\% & Home and Garden; Style &    13,978 &    0.8\% \\
        32 &  1.2\% & Travel                 &    10,440 &    0.6\% \\
        31 &  1.2\% & Technology; Business   &    23,283 &    1.3\% \\
        27 &  1.0\% &                        &    42,157 &    2.3\% \\
        25 &  0.9\% & Week in Review         &    17,107 &    0.9\% \\
        25 &  0.9\% & Home and Garden        &     5,546 &    0.3\% \\
        17 &  0.6\% & World; Washington      &    24,817 &    1.3\% \\
        17 &  0.6\% & Style; Magazine        &     1,519 &    0.1\% \\
    \bottomrule
    \end{tabular}
    \label{tab:newspaper_section}
\end{table*}

\begin{table}[]
    \centering
    \caption{Number of articles by newspaper category.}
    \begin{tabular}{@{}lr@{}}
\toprule
 articles   & 1,854,726 \\
 categories &    1,580 \\
\midrule
Business   &  291,982 \\
 Sports     &  160,888 \\
 Opinion    &  134,428 \\
 U.S.       &   89,389 \\
 Arts       &   88,460 \\
 World      &   79,786 \\
 Style      &   65,071 \\
 Obituaries &   19,430 \\
 Magazine   &   11,464 \\
 Travel     &   10,440 \\
\bottomrule    \end{tabular}
    \label{tab:newspaper_categories}
\end{table}

\subsection{Authors with a Thing for VA}

One next step was to see if there are certain authors with a thing for VA. If so, we could assume that at least some authors would use VA very consciously being well aware of its stylistic impact. Table~\ref{tab:authors} presents some authors with an inclination to use VA.

\begin{table*}[]
    \centering
        \caption{Ranking of VA-embracing authors (absolute number of VA per author -- percentage among all VA extracted -- total number of articles per author -- percentage among all articles in the corpus). Some author fields in the corpus were empty, so from the corpus itself we do not know for 15.5\% of extracted VA who authored them.}
    \begin{tabular}{@{}rrlrr@{}}
    \toprule
        VA &  2,646 & author                & articles & 1,854,726 \\
 \midrule
       411 & 15.5\% &                       &   961,052 &   51.8\% \\
        30 &  1.1\% & Holden, Stephen       &     5,098 &    0.3\% \\
        29 &  1.1\% & Maslin, Janet         &     2,874 &    0.2\% \\
        26 &  1.0\% & Vecsey, George        &     2,739 &    0.1\% \\
        23 &  0.9\% & Sandomir, Richard     &     3,140 &    0.2\% \\
        22 &  0.8\% & Ketcham, Diane        &      717 &    0.0\% \\
        20 &  0.8\% & Kisselgoff, Anna      &     2,661 &    0.1\% \\
        19 &  0.7\% & Dowd, Maureen         &     1,647 &    0.1\% \\
        19 &  0.7\% & Berkow, Ira           &     1,704 &    0.1\% \\
        18 &  0.7\% & Kimmelman, Michael    &     1,515 &    0.1\% \\
        17 &  0.6\% & Brown, Patricia Leigh &      568 &    0.0\% \\
        16 &  0.6\% & Pareles, Jon          &     4,090 &    0.2\% \\
        16 &  0.6\% & Chass, Murray         &     4,544 &    0.2\% \\
        15 &  0.6\% & Smith, Roberta        &     2,497 &    0.1\% \\
        15 &  0.6\% & Lipsyte, Robert       &      817 &    0.0\% \\
        15 &  0.6\% & Grimes, William       &     1,368 &    0.1\% \\
        15 &  0.6\% & Barron, James         &     2,188 &    0.1\% \\
        15 &  0.6\% & Anderson, Dave        &     2,735 &    0.1\% \\
        14 &  0.5\% & Stanley, Alessandra   &     1,437 &    0.1\% \\
        14 &  0.5\% & Haberman, Clyde       &     2,492 &    0.1\% \\
        \bottomrule
    \end{tabular}
    \label{tab:authors}
\end{table*}

\subsection{Anti-VA}

As we have shown in Table~\ref{tab:sources}, Michael Jordan is a universal super `source' for VA, a meme in its own right \citep{lovinger1999mike}. The \emph{Wall Street Journal} runs a website collecting all kinds of occurrences of ``Michael Jordans of \dots'', counting 1,505 occurrences in different media to date \citep{cohen2015michael}. On top of that, Barack Obama, in his honorific speech for Michael Jordan when awarding him the Medal of Freedom in November 2016, said the following: ``There is a reason you call someone the Michael Jordan of [something]. They know what you're talking about because Michael Jordan is the Michael Jordan of greatness. He is the definition of somebody so good at what they do that everybody recognizes it. That's pretty rare.'' \citep{payne2016obama}

Obama stretches VA past its breaking point by implementing recursivity, using an identical source and target (``Michael Jordan is the Michael Jordan of greatness''). Another way of doing that would be to use a non-modifying modifier, as has been acted out by German author Joachim Lottmann, who ironically referred to himself as ``the German version of Rainald Goetz'' \citep{seidl2004jugend}. Since Rainald Goetz, the `source' for this VA, is a German author himself, the whole formulation stops making sense as the modifier wasn't used to transfer the meaning into a different field, which lets this VA expression explode in irony. Another example for this sub-phenomenon will demonstrate that we are not talking about a single occurrence here: ``I am [right now] in Norway, the Sweden of Scandinavia,'' said Swiss satirist Gabriel Vetter in a recent interview, also leaving the modifier un-modified \citep{fopp2016wurst}.

We will conclude this parade of examples with a third class of anti-VA, quoting a passage which was returned when processing the NYT corpus: ``During their decade-long marriage, Elizabeth Taylor quipped that Burton was `the Frank Sinatra of Shakespeare.''' (NYT, 12 March 1989) The use of another human as modifier (here standing \emph{pars pro toto} for Shakespeare's oeuvre, too) is another attempt to wrap VA around itself. The same pattern is found in a well-known German example, Eckhard Henscheid's satirical opera guide ``Verdi ist der Mozart Wagners'' (1979). The description of these three ways to direct VA against itself was only possible because our exploratory large-scale analysis sharpened our senses for these kinds of absurd antonomasia.

\section{Summary and Outlook}

We started this project with the notion that Vossian Antonomasia is a stylistic device with a simple structure. It turned out, however, that it is a far richer phenomenon than we thought. Writing extraction rules for this phenomenon is anything but trivial. By resorting to Wikidata, we circumvented shortcomings of available NER tools. By discussing the most frequent VA extracted from the \emph{New York Times} corpus we added a quantifying approach to the discourse around paragons. We also presented first insights into the distribution of VA and found that certain journalistic genres have an inclination towards using this device, whereas others avoid it (``Sports'' and ``Arts'' sections vs. ``Business''). We also confirmed that VA is a language- and culture-dependent phenomenon, much like other stylistic devices \citep{koevecses2005metaphor}.

Our approach comes with known limitations. For one thing, we have focused our attention on just one grammatical variant of VA. Although our pragmatic limitation to the pattern ``the \dots of'' aimed at the most common shape of VA, we thereby missed expressions like ``the American Oscar Wilde'', ``Harlem's Mozart'', or ``the Japanese Nolan Ryan'' (all taken from the NYT corpus), which certainly leaves room for further work. The same can be said about our limitation to persons as source entities (living or dead, but in any case non-fictional and non-mythological). While working with our corpus we made some observations, though, that we don't want to withhold:
\begin{itemize}
    \item Productive `sources' for VA can be real persons as well as mythological or fictional characters (productive examples for the latter include Woody Allen's ``Zelig'', Cervantes' ``Don Quixote'' or James Hilton's ``Mr.~Chips'', but also sources like ``the [Santa Claus, Satan, Midas, Godzilla, King Kong, Cinderella, Pied Piper, Robin Hood, Energizer Bunny, Cassandra, Jupiter, Icarus] of --''.
\item Animals can be targets, even if the source is a human, like in ``\emph{Sea Hero} is the Bobo Holloman of racing'' (NYT, 1993) or ``\emph{Bonfire}, the Michael Jordan of dressage horses'' (NYT, 1998).
\item Personifications can be part of a VA expression, that is, the use of individual persons/characters as sources for companies, clubs, bands or places as `targets', like in ``\emph{Sturm, Ruger} is the \emph{Benedict Arnold} of the gun industry'' (NYT, 1989), ``\emph{Aerosmith}, the \emph{Dorian Gray} of rock bands'' (NYT, 1993), ``the \emph{Hudson} has been the \emph{John Barrymore} of rivers, noble in profile but a sorry wreck'' (NYT, 1996), ``the \emph{National Collegiate Athletic Association}, the \emph{Kenneth Starr} of sports'' (NYT, 1998).
\end{itemize}

In conclusion, we would like to propose more holistic approaches to the phenomenon rather than just building on Lakoff's `paragons', that is, the `sources'. With quantitative explorations of Vossian Antonomasia like the one presented here, we can capture the phenomenon as a whole, including source, target and (if provided) modifier. We will find a network of people, connected by a diversity of modifiers, with nodes that can be sources or targets (or both, as shown by the Obama example), a network that can help us understand the hidden patterns of role models and how they differ from country to country, from language to language.

\section{Acknowledgement}
We would like to thank the Information School at the University of Sheffield for granting one of the authors (F.\,F.) a visiting scholarship in spring 2017 which helped to finish our research.

\bibliography{references}

\begin{thebibliography}{16}
\expandafter\ifx\csname natexlab\endcsname\relax\def\natexlab#1{#1}\fi

\bibitem[{Bergien(2013)}]{bergien2013names}
Angelika Bergien. 2013.
\newblock \href
  {https://onomasticafelecan.ro/iconn2/proceedings/1_01_Bergien_Angelika_ICONN_2.pdf}
  {Names as frames in current-day media discourse}.
\newblock In \emph{Name and Naming. Proceedings of the second international
  conference on onomastics}, pages 19--27, Cluj-Napoca. Editura Mega.

\bibitem[{Bird et~al.(2009)Bird, Klein, and Loper}]{bird2009nltk}
Steven Bird, Ewan Klein, and Edward Loper. 2009.
\newblock \emph{Natural Language Processing with Python}, 1st edition.
\newblock O'Reilly Media, Inc.

\bibitem[{Cohen et~al.(2015)Cohen, Foster, Oshinsky, Keegan, and
  McGinty}]{cohen2015michael}
Ben Cohen, Geoff Foster, Matthew Oshinsky, Jon Keegan, and Tom McGinty. 2015.
\newblock \href {http://graphics.wsj.com/michael-jordan-of/} {The {Michael
  Jordan} of \dots}.
\newblock Website.

\bibitem[{Fischer and Wälzholz(2014)}]{fischer2014napoleon}
Frank Fischer and Joseph Wälzholz. 2014.
\newblock \href
  {https://www.umblaetterer.de/wp-content/uploads/2014/12/vossanto_fas.png}
  {{Jeder kann Napoleon sein. Vossianische Antonomasie: Eine Stilkunde}}.
\newblock \emph{Frankfurter Allgemeine Sonntagszeitung}, page~34.
\newblock Dec 21, 2014.

\bibitem[{Fopp(2016)}]{fopp2016wurst}
Andrea Fopp. 2016.
\newblock \href {http://www.tageswoche.ch/de/2016_36/leben/728705/} {{``Die
  Wurst ist Geheimnis, ist Gefahr, ist Geisterbahn'' [interview with Gabriel
  Vetter]}}.
\newblock \emph{TagesWoche}.
\newblock Sep 1, 2016.

\bibitem[{Grgić and Nikolić(2013)}]{grgic2013cowboys}
Ana Grgić and Davor Nikolić. 2013.
\newblock \href {https://bib.irb.hr/prikazi-rad?rad=651849} {{The Cowboys, the
  Poets, the Professor} \dots -- {Antonomasia in Croatian Sports Discourse}}.
\newblock In G.~Kišiček and I.Ž. Žagar, editors, \emph{What Do We Know
  about the World? Rhetorical and Argumentative Perspectives}, pages 408--428.
  University of Windsor; Pedagoški inštitut, Windsor; Ljubljana.

\bibitem[{Holmqvist and Płuciennik(2010)}]{holmqvist2010princess}
Kenneth Holmqvist and Jarosław Płuciennik. 2010.
\newblock \href {https://doi.org/10.1515/9783110230215} {{Princess Antonomasia
  and the Truth: Two Types of Metonymic Relations}}.
\newblock In Armin Burkhardt and Brigitte Nerlich, editors, \emph{Tropical
  Truth(s). The Epistemology of Metaphor and Other Tropes}, pages 373--381. De
  Gruyter, Berlin/New York.

\bibitem[{Jäschke et~al.(2017)Jäschke, Strötgen, Krotova, and
  Fischer}]{jaeschke2017nutella}
Robert Jäschke, Jannik Strötgen, Elena Krotova, and Frank Fischer. 2017.
\newblock \href
  {http://www.dhd2017.ch/wp-content/uploads/2017/03/Abstractband_def3_M\%C3\%A4rz.pdf\#page=122}
  {{``Der Helmut Kohl unter den Brotaufstrichen''. Zur Extraktion Vossianischer
  Antonomasien aus großen Zeitungskorpora}}.
\newblock In \emph{Proceedings of DHd 2017}, DHd '17, pages 120--124. Digital
  Humanities im deutschsprachigen Raum.

\bibitem[{Kövecses(2005)}]{koevecses2005metaphor}
Zoltán Kövecses. 2005.
\newblock \emph{Metaphor in Culture: Universality and Variation}.
\newblock Cambridge University Press, Cambridge.

\bibitem[{Lakoff(1987)}]{lakoff1987women}
George Lakoff. 1987.
\newblock \emph{Women, Fire, and Dangerous Things. What Categories Reveal about
  the Mind}.
\newblock The University of Chicago Press, Chicago.

\bibitem[{Lausberg(1960)}]{lausberg1960handbuch}
Heinrich Lausberg. 1960.
\newblock \emph{Handbuch der literarischen Rhetorik. Eine Grundlegung der
  Literaturwissenschaft}.
\newblock Hueber, München.

\bibitem[{Lovinger(1999)}]{lovinger1999mike}
Caitlin Lovinger. 1999.
\newblock \href {https://nyti.ms/2scmu3i} {{They're Sort of Like Mike}}.
\newblock \emph{New York Times}.
\newblock Jan 17, 1999.

\bibitem[{Payne(2016)}]{payne2016obama}
Marissa Payne. 2016.
\newblock \href {http://wapo.st/2fD96iF} {Obama tells {Michael Jordan} he's
  `more than just an {Internet} meme,' then makes {MJ} tear up}.
\newblock \emph{Washington Post}.
\newblock Nov 22, 2016.

\bibitem[{Sandhaus(2008)}]{sandhaus2008nyt}
Evan Sandhaus. 2008.
\newblock \href {https://catalog.ldc.upenn.edu/LDC2008T19} {{The New York Times
  Annotated Corpus LDC2008T19}}.
\newblock {DVD}, Linguistic Data Consortium, Philadelphia.

\bibitem[{Seidl(2004)}]{seidl2004jugend}
Claudius Seidl. 2004.
\newblock {Jugend ohne Sex. Ein Manifest gegen die Wirklichkeit: Joachim
  Lottmanns Roman ``Die Jugend von heute''}.
\newblock \emph{Frankfurter Allgemeine Sonntagszeitung}, page~34.
\newblock Oct 17, 2004.

\bibitem[{Vrande\v{c}i\'{c} and Kr\"{o}tzsch(2014)}]{vrandecic2014wikidata}
Denny Vrande\v{c}i\'{c} and Markus Kr\"{o}tzsch. 2014.
\newblock \href {https://doi.org/10.1145/2629489} {{Wikidata: A Free
  Collaborative Knowledgebase}}.
\newblock \emph{Communications of the ACM}, 57(10):78--85.

\end{thebibliography}
\bibliographystyle{acl_natbib}

\end{document}